\documentclass[a4paper,11pt]{article}
\usepackage{times}
\usepackage{amsmath,amssymb,amsfonts,amsthm}
\usepackage{graphicx}
\usepackage[colorlinks=true,urlcolor=blue]{hyperref}
\def\be{\begin{equation}}
\def\ee{\end{equation}}
\def\bea{\begin{eqnarray}}
\def\eea{\end{eqnarray}}
\def\lb{\label}
\def\ct{\cite}
\def\bi{\bibitem}
\def\vec#1{\mbox{\boldmath$#1$}}
\def\lgth{[\,\mbox{length}\,]}
\def\gam{\gamma}

\def\sig{\sigma}
\def\om{\omega}
\def\Om{\Omega}
\def\ptl{\partial}

\def\cqg{{\em Class. Quantum Grav.\/} }
\def\grg{{\em Gen. Rel. Grav.\/} }
\def\jcap{{\em J. Cosmol. Astropart. Phys.\/} }
\def\jmp{{\em J. Math. Phys.\/} }
\def\mn{{\em Mon. Not. R. Astron. Soc.\/} }
\def\prd{{\em Phys. Rev.\/} D }
\def\prl{{\em Phys. Rev. Lett.\/} }

\begin{document}
\title{\sc Geometrical order--of--magnitude estimates for spatial
curvature in \\ realistic models of the Universe}
\author{{\sc Thomas Buchert}${}^{1}$\thanks{E--mail: {\tt
      buchert@obs.univ-lyon1.fr}},
{\sc $\;$George F.R. Ellis}${}^{2}$\thanks{E--mail: {\tt
      George.Ellis@uct.ac.za}} $\;$and
{\sc Henk van Elst}${}^{3}$\thanks{E--mail: {\tt
      hvanelst@merkur-fh.org}} \\
{\small\em ${}^{1}$Universit\'{e} Lyon 1, Centre de Recherche
Astrophysique de Lyon, 9 avenue Charles Andr\'{e}} \\
{\small\em 69230 Saint--Genis--Laval, France} \\
{\small\em ${}^{2}$Cosmology \& Gravity Group, Department of
Mathematics and Applied Mathematics, University of Cape Town} \\
{\small\em Rondebosch 7701, South Africa} \\
{\small\em ${}^{3}$Fakult\"{a}t I: Betriebswirtschaft, Merkur
Internationale FH Karlsruhe, Karlstra\ss e 36--38} \\
{\small\em 76133 Karlsruhe, Germany}}

\date{\normalsize{April 28, 2009}}
\maketitle
\begin{abstract}
The thoughts expressed in this article are based on remarks made
by J\"{u}rgen Ehlers at the Albert--Einstein--Institut,
Golm, Germany in July 2007. The main objective of this article is
to demonstrate, in terms of plausible order--of--magnitude
estimates for geometrical scalars, the relevance of spatial
curvature in realistic models of the Universe that describe the
dynamics of structure formation since the epoch of
matter--radiation decoupling. We introduce these estimates with a
commentary on the use of a quasi--Newtonian metric form  in this
context.

\end{abstract}
\begin{flushleft}
PACS number(s): 04.20.-q, 04.20.Cv, 98.80.-k, 98.80.Jk
\\
Preprint number(s):
\href{http://arxiv.org/abs/0906.0134}{arXiv:0906.0134 [gr-qc]}
\hfill
DOI:
\href{http://dx.doi.org/10.1007/s10714-009-0828-4}{10.1007/s10714-009-0828-4}
\end{flushleft}
\begin{flushright}
{\large\em In Memoriam:\/} \,\, {\large\em J\"{u}rgen Ehlers\/}\
{\em (1929--2008)\/}
\end{flushright}

\section{Introduction}
\lb{sec:intro}
In July 2007, J\"{u}rgen Ehlers gave a talk at the
Albert--Einstein--Institut, Golm, Germany on the relevance of
spatial curvature in models of structure formation in the Universe
since the epoch of matter--radiation decoupling. This article aims
to make his comments publicly available, after providing a
contextual setting by first commenting on the more usual approach to
these issues in structure formation studies.

The so--called {\em longitudinal gauge\/} is often employed in the
study of scalar perturbations at a Friedmannian background
cosmology, and is considered a preferred frame because it offers an
explicit Newtonian limit; cf. Ref.~\ct{wal1984}.\footnote{This limit
obtains the standard Eulerian formulation of the equations of
Newtonian cosmology. An alternative Lagrangian formulation of these
equations is presented in Ref.~\ct{ehlbuc1997}. In the Lagrangian
representation of the equations of relativistic cosmology the
natural Newtonian limit is obtained in the matter--comoving frame;
cf. Sec.~4.2.1 of Ref.~\ct{buc2008}.} While its local foundations
are unambiguous (see, e.g., Refs.
\ct{bar1980,muketal1992,kodamasasaki}), its global use is less
clear; but its status is such that it has been elevated  to a
paradigm: {\em the dynamics of the inhomogeneous, real Universe can
be described globally, from the largest scales down to the scales
where spatial curvature effects become significant, by a single
quasi--Newtonian metric form\/}. This is most lively reflected in
articles that deal with the possible impact of inhomogeneities on
expansion properties of the Universe (the so--called ``backreaction
problem''), a topic that --- especially recently --- is often
discussed in the quasi--Newtonian setting (see, e.g., Refs.
\ct{futamase1,futamase2,abramo1,branden1,parry} and many others).

In the {\em longitudinal gauge\/}, fixed in relation with a
$3+1$ decomposition of the cosmological spacetime manifold
according to Arnowitt, Deser and Misner~\ct{adm1962}, the lapse
function and the spatial metric are specified so as to
provide a ``perturbed Newtonian setting''. The metric form for
the physical spacetime is set to be
\be
\lb{postnewton1}
{}^{4}\!{\bf g} = -\,N^{2}(t,x^{k})\,{\rm d}t\otimes{\rm d}t
+ g_{ij}(t,x^{k})\,{\rm d}x^{i} \otimes {\rm d}x^{j} \ ,
\ee
where the lapse function $N$ and the spatial metric coefficients
$g_{ij}$ of a family of spacelike 3--surfaces
${\cal S}:\{t=\mbox{constant}\}$ orthogonal to
an irrotational and shearfree timelike reference congruence
$\vec{n} = N^{-1}\,\ptl_{t}$ are given by
\be
\lb{postnewton2}
N^{2} = \ell_{0}^{2}a^{2}(t)[1+2\Phi(t,x^{k})] \ ,
\hspace{10mm}
g_{ij} = \ell_{0}^{2}a^{2}(t)[1-2\Psi(t,x^{k})]\,
\gamma_{ij} \ ,
\ee
implying the vanishing of each of the spatial Cotton--York tensor,
$C_{ij}(g)=0$, and the magnetic Weyl curvature,
$H_{ab}(\vec{n})=0$, and so ensuring the strict absence of
gravitational radiation (cf. Ref.~\ct{hveell1998}). Here, $a(t)$
denotes the dimensionless scale factor of a spatially
homogeneous and isotropic solution of Einstein's field equations,
contained in the metric form (\ref{postnewton2}) for $0 = \Phi
= \Psi$, and $\gamma_{ij}$ is a spatial metric of constant
curvature, i.e., $R(\gam)=\mbox{constant}$. Backreaction effects
are not taken into account. Frequently, the simplifying choice
\be
\lb{postnewton3}
\gam_{ij} = \delta_{ij}
\ee
(so that $R(\gam)=0$ holds true) is made. The constant $\ell_{0}$
represents the unit of the physical dimension $\lgth$, $t$ is the
dimensionless (conformal) local coordinate time, and the
dimensionless $x^{i}$ (being local coordinates in the tangent
spaces at any spatial position in the 3--surfaces in standard
general relativity) are here considered as ``background
coordinates'', i.e., the inhomogeneous metric perturbations
encoded in the functions $\Phi$ and $\Psi$ are, in the framework
of gauge--invariant cosmological perturbation theory, considered
as functions of {\em globally defined coordinates\/}. In this
framework, $\Phi$ and $\Psi$ correspond to Bardeen's
gauge--invariant potentials for scalar metric
perturbations~\ct{bar1980}. For a perfect fluid energy--momentum
tensor, upon neglecting terms quadratic in peculiar velocities,
they can be set equal to each other:
\be
\lb{postnewton4}
\Phi = \Psi
\ee
(cf. Ref.~\ct[p~223]{muketal1992}).

The above metric setting comes with a list of restrictions
that have to be imposed on the perturbation function
$\Psi (t,x^{i})$; in the standard literature, this list
comprises the following conditions:
\be
\lb{restrictions}
|\Psi| \ll 1 \ , \quad
\left|\frac{\ptl\Psi}{\ptl t}\right|^{2} \ll
\frac{1}{a^{2}}\gam^{ij}D_{i}\Psi D_{j}\Psi \ , \quad
(\gam^{ij}D_{i}\Psi D_{j}\Psi)^{2} \ll
\gam^{ik}\gam^{jl}D_{i}D_{j}\Psi D_{k}D_{l}\Psi \ ;
\ee
the operator $D_{i}$ denotes the covariant derivative associated
with the constant curvature spatial metric $\gam_{ij}$. Note that
the last inequality, which is a necessary condition for
perturbations in the spatial metric to be small, only {\em
compares\/} the sizes of two specific spatial curvature terms; it
does not say that spatial curvature has to be small per se.

Among the many careful papers that do include the above list, we
select the recent paper by Ishibashi and Wald~\ct{ishwal2006}. On
p~238 of their work these authors assert that ``the metric
(\ref{postnewton1})--(\ref{postnewton3}) appears to very accurately
describe our Universe on {\em all scales\/}, except in the immediate
vicinity of black holes and neutron stars'', and they continue:
``The basis for this assertion is simply that the FLRW metric
appears to provide a very accurate description of all phenomena
observed on large scales, whereas Newtonian gravity appears to
provide an accurate description of all phenomena observed on small
scales.''

They also sketch a typical cosmologically relevant energy--momentum
tensor that assumes the form of a {\em homogeneous\/} perfect fluid
for ``homogeneously distributed matter'' and an inhomogeneous
continuum of ``dust'' that, this latter, could be approximated as
\be
\label{dustmatter}
{\bf T}^{(m)} \approx \rho(t,x^{i})\,
{\rm d}t\otimes{\rm d}t \ ,
\ee
on the assumption of small (non--relativistic) peculiar velocities.
In general, in cosmological perturbation theory, peculiar velocities
are being taken into account, e.g., to first order of smallness.

We take this paper as an example of the aforementioned
paradigm. We do not fully enter the issue raised in it
related to the recent discussion on whether backreaction
effects may account for the dark energy problem of the
standard model of cosmology or not (we refer the reader to Ref.
\ct{koletal2006}, and especially to the recent paper
\ct{koletal2008}, where it is also commented on the applicability
of the quasi--Newtonian metric form, and the review papers
\ct{ellbuc2005,rasanen,buc2008}).

However, let us add a remark that shows that it is, from a
fully relativistic point of view, implausible that the
physical spacetime metric in the form of
Eqs.~(\ref{postnewton1})--(\ref{postnewton3}) can be carried to
the point of describing a realistic model of the Universe, if
we employ this metric as an approximate solution to Einstein's
field equations. While it is true that this metric can provide
a good local description of either a perturbed expanding
Universe or a quasi--static domain of local matter condensations,
it is not at all clear to what extent it can represent both
{\em simultaneously\/}. The specific issue is: ``How large a
domain in space and time can be covered by such a `global'
coordinate system in a realistic model representing both the
dynamical expanding Universe and imbedded local large--scale
voids?'' This is a crucial issue in cosmology.

The strong conclusions that are advanced, e.g., in the paper by
Ishibashi and Wald~\ct{ishwal2006}, can, of course, be tested on the
grounds of a realistic evaluation of this metric ansatz as an
approximate solution to the field equations of general relativity,
irrespective of a particular framework from which this metric ansatz
has been derived. We point out here that this metric form and the
accompanying list of restrictions does not contain any condition
that forces us to choose a particular 4--velocity field, say,
$\vec{u}$ of a fluid continuum evolving in this spacetime, except it
is implicit that peculiar velocities will be small. Potential
danger is associated with the form of the energy--momentum
tensor~(\ref{dustmatter}) for an inhomogeneous dust continuum,
taken from Ref.~\ct[Eq.~(4)]{ishwal2006}: if, in
the chosen time slicing of the cosmological spacetime manifold,
we take this form of the energy--momentum tensor literally, i.e.,
ignore the approximation sign which is supposed to imply
quantitatively negligible peculiar velocities, then there is
strictly {\em no relation of the quasi--Newtonian metric form to
inhomogeneities\/} (cf. Ref.~\ct[p~3566]{hveell1998}) and, e.g.,
questions such as the backreaction of inhomogeneities {\em cannot
even be addressed\/}. We would simply be looking at a spatially
homogeneous solution in an odd coordinate system, which makes
the metric ``look'' spatially inhomogeneous in terms of the
perturbation function $\Psi(t,x^{i})$.

Indeed, if we choose a fluid 4--velocity
field $\vec{u}$ normal to the spacelike 3--surfaces defined by
this metric form (which is equivalent to setting peculiar
velocities exactly to zero), then this implies a {\em shearfree
fluid motion\/} (see, e.g., Ref.~\ct{hveell1998}). It then follows
that the fluid in this spacetime must be spatially homogeneous
``in most cases''. More precisely, if the energy--momentum tensor
only represents a dust matter source, then shearfree solutions
always describe a spatially homogeneous continuum; see Theorem 1
and Corollary 1 on p~1210 in the paper by Collins and
Wainwright \ct{colwai1983}. In the case of perfect fluid sources,
there are some inhomogeneous solutions that are, however, of no
obvious cosmological relevance; see, e.g.,
Refs.~\ct{colwai1983,colwhi1984,whicol1984}. Already this remark
makes clear that a thoughtless application of the quasi--Newtonian
metric form can quickly run into trouble: when the approximation made
ignores the peculiar velocity terms in the field equations, it is
in danger of running into effects related to these restrictions
applying to the corresponding exact solutions.

A warning against the assumption that a Newtonian limit of this
kind is without problems in the cosmological context is the
following: it is an exact theorem that shearfree dust solutions
of Einstein's field equations cannot both expand and rotate, i.e.,
\be
\lb{eq:SF}
\sigma=0 \ , \quad p=0 \quad\Rightarrow\quad \theta\om = 0 \ ;
\ee
see Ref.~\ct{ell1967}. However, shearfree solutions of the
corresponding Newtonian equations do exist where this is {\em not\/}
true: they can both expand and rotate; cf. Ref.~\ct{nar1963}.
Consequently, the Newtonian limit is singular. Consider a sequence
$GRT(i)_{\sig=0}$ of relativistic shearfree dust solutions with a
limiting solution $GRT(0)_{\sig=0}$ that constitutes the Newtonian
limit of this sequence. The latter solution will necessarily satisfy
Eqs.~(\ref{eq:SF}) because every solution $GRT(i)_{\sig=0}$ in the
sequence does so. The corresponding exact Newtonian solution
$NGT(0)_{\sig=0}$ will therefore also necessarily satisfy
Eqs.~(\ref{eq:SF}). But the Newtonian solutions $NGT(j)_{\sig=0}$
that do not satisfy Eqs.~(\ref{eq:SF}) are clearly not obtainable as
limits of any sequence of relativistic solutions $GRT(j)_{\sig=0}$.
Assuming Einstein's field equations represent the genuine theory of
gravitational interactions in the physical Universe, with solutions
of the Newtonian equations an acceptable approximation to
relativistic solutions under suitable circumstances, this result
tells us that not all Newtonian solutions are indeed such
acceptable approximations.

\section{Notes of J\"{u}rgen Ehlers' remarks}
We wish to emphasise that the subsequent notes shall not be
understood as a ``photographic reproduction'' of J\"{u}rgen Ehlers'
original blackboard writings, given during a talk on Thu, July 26,
2007 when two of us (TB and HvE) were present, but rather that the
essence of his remarks is being truthfully reproduced and
summarised.

The starting point of J\"urgen's considerations for gravitating
physical systems was the central assumption that Einstein's general
theory of relativity constitutes the ``correct'' theory of
gravitational interactions on the scales of the solar system,
stars, neutron stars and black holes, on which this theory has been
reliably tested to remarkable accuracy. It is our recollection
that J\"urgen was very cautious here and specifically related his
comments to tested scales only. However, at the end of his talk
he pointed out that an extrapolation to cosmological scales of the
matter he had raised was conceivable and so could be a natural
related investigation. The intention of the argument he gave was
to illustrate the fact that, even if metrical perturbations for a
gravitating system of the above mentioned scales are small
in magnitude, the {\em derivatives\/} of such perturbations, the
second--order ones in particular, can be physically significant.

\subsection{Metric level}
We henceforth consider a domain ${\cal D}$ of a given spacetime
manifold ${\cal M}$. On ${\cal D}$ we decompose the physical
spacetime metric ${}^{4}\!{\bf g}$ into a leading--order term
$\stackrel{0}{{\bf g}}$ and a term of small relative deviations
$\boldsymbol{h}$, the latter referred to as perturbations
in ${}^{4}\!{\bf g}$:
\bea
\lb{4metric}
{}^{4}\!{\bf g}
& = & \underbrace{\stackrel{0}{{\bf g}}}_{{\rm scale\ of
\ changes}:\ D} + \underbrace{\boldsymbol{h}}_{{\rm scale\ of
\ changes}:\ d} \nonumber \\
& \approx & {\cal O}(1) + {\cal O}(\varepsilon) \ ;
\eea
$\boldsymbol{h}$ is assumed to be of first order in an
appropriate dimensionless smallness parameter $\varepsilon$
with $|\varepsilon| \ll 1$.\footnote{For reasons of
notational ease we will subsequently drop the superscript
``4'' from ${}^{4}\!{\bf g}$.} With appreciable changes
experienced in $\stackrel{0}{{\bf g}}$ we associate a
macrosopic characteristic spacetime scale $D$, while,
analogously, with appreciable changes in $\boldsymbol{h}$
we associate a microsopic characteristic spacetime scale
$d$. The {\em scale ratio\/}
\be
\frac{D}{d}
\ee
thus constitutes a dimensionless physical quantity of special
interest for order--of--magnitude estimates in respect to
leading--order physical effects in two--scale gravitating
systems of the kind outlined.

\subsection{Connection level}
At the level of the spacetime connection the decomposition of
Eq.~(\ref{4metric}) suggests that
\bea
\lb{conest}
\boldsymbol{\Gamma}({\bf g})
& = & {\bf g}^{-1}\boldsymbol{\partial}{\bf g}
\nonumber \\
& \approx & (\stackrel{0}{{\bf g}}{}^{-1}+\boldsymbol{h})
(\boldsymbol{\partial}\!\stackrel{0}{\bf g}
+\boldsymbol{\partial}\boldsymbol{h})
\nonumber \\
& \approx & \left(1+\varepsilon\right)\frac{1}{D}
\left(1+\varepsilon\frac{D}{d}\right)
\nonumber \\
& \approx & \frac{1}{D}\left(1+\varepsilon\frac{D}{d}
+\ldots\right)
\ \approx \ \stackrel{0}{\boldsymbol{\Gamma}}
+ \stackrel{1}{\boldsymbol{\Gamma}} \ .
\eea
%

\subsection{Curvature level}
Einstein's field equations of gravitational interactions are
formulated at the level of spacetime curvature. It is at
{\em this\/} level that we notice/observe the characteristic
dynamical features of relativistic gravitational physics. For the
spacetime curvature we obtain
\bea
\lb{curvest}
\boldsymbol{R}({\bf g})
& = & \boldsymbol{\partial\Gamma}
+ \boldsymbol{\Gamma}\boldsymbol{\Gamma}
\ = \ {\bf g}^{-1}\boldsymbol{\partial}^{2}{\bf g}
+ ({\bf g}^{-1}\boldsymbol{\partial}{\bf g})^{2}
\nonumber \\
& \approx & \left(1+\varepsilon\right)\frac{1}{D^{2}}
\left[\,1+\varepsilon\left(\frac{D}{d}\right)^{2}\,\right]
+ \left(1+\varepsilon\right)^{2}\frac{1}{D^{2}}
\left(1+\varepsilon\frac{D}{d}\right)^{2}
\nonumber \\
& \approx & \frac{1}{D^{2}}\left[\,1
+ \varepsilon\left(\frac{D}{d}\right)^{2} + \ldots\,\right]
\ \approx \ \stackrel{0}{\boldsymbol{R}}
+ \stackrel{1}{\boldsymbol{R}} \ .
\eea
We note that, even when deviations in the metrical amplitude
itself are small, resultant deviations in the spacetime curvature
may become significant, depending on the specific value of
the scale ratio $D/d$. A significant influence on the
spacetime curvature will therefore arise provided that
\be
\varepsilon\left(\frac{D}{d}\right)^{2} \approx {\cal O}(1) \ ,
\ee
or, in a $\log(D/d)$--$\log(\varepsilon)$ representation,
whenever
\be
\log\left(\frac{D}{d}\right)
\approx -\,\frac{1}{2}\,\log(\varepsilon) \ .
\ee
We conclude that the application of a
strictly Newtonian or quasi--Newtonian discription of
gravitational interactions to gravitating systems with two
characteristic spacetime scales becomes justified when for this
system at least the two constraints (i)~$\varepsilon \ll 1$ and
(ii)~$\varepsilon(D/d)^{2} \ll 1$ are simultaneously satisfied.
Order--of--magnitude considerations of this kind are relevant for
all gravitating systems for which characteristic spacetime
scales $D$ and $d$ can be identified.

J\"urgen, confining himself to spatial considerations within
the scheme outlined above, presented a simple numerical example
for the solar system, where general relativity is well
established. He chose $\varepsilon \approx 4.23\times 10^{-6}$
for the magnitude of deviations in the spatial metric from a
Euclidian geometry, $D \approx 150 \times 10^{6}\,{\rm km}$ for
the radius of the Earth's orbit around the Sun, and
$d \approx 6.95\times 10^{5}\,{\rm km}$ for the radius of the
Sun to find $\varepsilon(D/d)^{2} \approx 0.20$. J\"urgen remarked
that analogous order--of--magnitude estimates are helpful
indicators for the relevance of curvature effects, too, in
gravitating systems of the scales of galaxies, clusters of
galaxies, and the entire Universe itself. Taking the liberty
to extrapolate the validity of general relativity (unmodified)
all the way up to the scales of the observable Universe (and
thus ignoring the subtleties and intricacies associated with
an averaging approach to cosmology which would be more
appropriate here; cf. Refs.~\ct{ell1984}, \ct{buc2008} or
\ct{wil2007}), we now turn to following J\"urgen's suggestion
and work out the numerical details for the astrophysical and
cosmological systems that he had mentioned.

\section{Numerical examples}
\lb{sec:numex}
We will adapt J\"urgen's argument to spatial geometries and
estimate the order--of--magnitude of perturbations in the spatial
metric and the spatial curvature in the context of structures
observed in the matter distribution at different scales up to
the scale of the observable Universe (see also
Refs.~\ct{ellbuc2005} and \ct{buccar2008}); the required reduction
of 4--D considerations to 3--D considerations can be obtained
in terms of a ``thin sandwich approach'' as employed, e.g., in the
appendix of Ref.~\ct{buc2000}. To this end we
project typical spacetime length scales for respective
astrophysical and cosmological gravitating systems into the
present--day spacelike 3--surface ${\cal S}_{0}:\{t=t_{0}\}$ of
observers located on Earth; for these observers (which are in
relative motion to the cosmological Hubble flow) the length scales
we thus consider constitute instantaneous proper spatial distances
(see Refs.~\ct{ehl1961} or \ct{ellhve1999} for technical details).
Here we choose the macroscopic scale $D$ to represent the
{\em size\/} of the gravitating system, which we take to be its
physical diameter, while we choose the microscopic scale $d$ to
represent the {\em smoothing scale\/}, i.e., the scale below which
we neglect the influence of inhomogeneities in the matter
distribution and the geometry. In general $d$ will be the physical
diameter of a gravitating substructure of the system in question,
with $D/d \gg 1$. The working hypothesis of our investigations
shall be the frequently encountered assumption that overall the
geometry of space within gravitating systems at astrophysical and
cosmological scales is flat and that deviations from the Euclidian
geometry can be modelled in terms of small perturbations. We
emphasise that the idea is to see what realistic order--of--magnitude
estimates can be obtained for the contribution of representative
structures in the distribution of matter to the curvature of
space at different scales.

For a specific two--scale gravitating system, we will evaluate
an order--of--magnitude estimate for the following two
characteristic dimensionless parameters:
\begin{itemize}
\item[(i)] the ratio between the Schwarzschild and
the proper physical radii of the system, i.e.,
\be
\varepsilon := \frac{4GM}{c^{2}D} \ ,
\ee
(note that in the cosmologial context, where $M$ typically
scales as $D^{3}$, cf.\ Sec.~12.1 of Ref.~\ct{sch1988}, the
parameter $\varepsilon$ scales as $D^{2}$), and
\item[(ii)] the spatial curvature perturbation
\be
\varepsilon\left(\frac{D}{d}\right)^{2} \ .
\ee
\end{itemize}

Besides the Earth's orbit around the Sun (A1), we select as
further representative gravitating systems at astrophysical
scales a typical galaxy (A2) and a typical cluster of
galaxies (A3). A galaxy we assume to contain 100 billion solar
mass stars within a spatial domain of diameter
$100000~{\rm ly}$, while we model a cluster of galaxies as
containing 1000 galaxies (of 100 billion solar mass stars each)
within a spatial domain of diameter $5~{\rm Mpc}$.\footnote{A
nice starting point for obtaining realistic values for the
masses $M$ and diameters $D$ and $d$ is the convenient online
encyclopedia \href{http://en.wikipedia.org}{en.wikipedia.org}.
See also the comprehensive summary by Cox~\ct{cox2007}.} In
both of these cases, we are deliberately confining ourselves
to considerations of luminous (baryonic) matter only. The
effects of a considerable factor (possibly 10, or larger still)
due to additional components of non--luminous matter in the mass
content $M$ on the parameters $\varepsilon$ and
$\varepsilon(D/d)^{2}$ can be easily traced.

For the examples C1 to C3 of gravitating systems at
cosmological scales, we obtain the information on the mass
content $M$ of a sphere of radius $D/2$ in Euclidian space
from the value
\be
\lb{wmapdens}
\rho_{m} \approx 2.58\times 10^{-27}~{\rm kg}~{\rm m}^{-3}
\ee
for an average (baryonic and dark matter) mass density,
which derives from the WMAP five--year data results
$\Om_{m}h^{2} \approx 0.14$ and $h \approx 0.71$ (so
$\Om_{m} \approx 0.27$) taken from the work by Hinshaw
{\em et al\/}~\ct{hinetal2009}. In the cosmic void example C1
we assume that the underdensity is balanced by a
corresponding overdensity associated with the enveloping
wall structure so that the value of $\rho_{m}$ in
Eq.~(\ref{wmapdens}) can be employed without modification.
Conservative values for the scales $D$ and $d$ were selected
for examples C1 and C2, which represent typical
cases of large--scale structures. We consider the particular
choice of $300h^{-1}~{\rm Mpc}$ for the scale of statistical
homogeneity in the observable Universe as a conservative lower
limit.\footnote{The  determination of this scale depends on
the statistical measure of inhomogeneity employed. Often a
lower value is quoted that is, however, based on weak 
statistical measures like the two--point correlation function.
Inhomogeneities mirrored by morphological  differences show
up in higher--order correlations of the distribution that
can be captured by calulating the Minkowski Functionals; see,
e.g., Kerscher {\em et al\/}~\ct{keretal2001}.} By definition
there are no matter structures beyond this scale. Therefore,
in the example C3 of the present--day Hubble sphere we only
evaluate the parameter $\varepsilon$. It turns out that with
$D=2r_{H_{0}}=2c/H_{0}$ and $M$ computed from $\rho_{m}$
of Eq.~(\ref{wmapdens}), in this case $\varepsilon$ becomes
identical to the matter density parameter $\Om_{m}$. We
ask the reader to choose their own set of realistic values
for $M$, $D$ and $d$ in order to obtain further estimates
for $\varepsilon$ and $\varepsilon(D/d)^{2}$.
\begin{table}[!htb]
  \centering
    \begin{tabular}{c|c|c|c|c|c}
      \hline\hline
      Gravitating system / & Mass & Diameters & & & \\
      Smoothing scale & $M$ & $D$ and $d$ & $D/d$ & $\varepsilon$ & $\varepsilon(D/d)^{2}$ \\
      \hline\hline
      A1: Earth's orbit / & $\approx M_{\odot}$ & $300\times 10^{6}~{\rm km}$ & & & \\
      Sun & ($1.99\times 10^{30}~{\rm kg}$) & $1.39\times 10^{6}~{\rm km}$ & $216$ & $4.24\times 10^{-6}$ & $0.20$ \\
      \hline
      A2: Galaxy / & $\approx 10^{11}~M_{\odot}$ & $100000~{\rm ly}$ & & & \\
      Open star cluster & ($1.99\times 10^{41}~{\rm kg}$) & $30~{\rm ly}$ & $3333$ & $6.23\times 10^{-7}$ & $6.92$ \\
      \hline
      A3: Cluster of galaxies / & $\approx 10^{14}~M_{\odot}$ & $5~{\rm Mpc}$ & & & \\
      Galaxy & ($1.99\times 10^{44}~{\rm kg}$) & $0.03~{\rm Mpc}$ & $167$ & $3.82\times 10^{-6}$ & $0.11$ \\
      \hline
      C1: Void / & $\approx (1/6)\pi\rho_{m}D^{3}$ & $30h^{-1}~{\rm Mpc}$ & & & \\
      Wall & ($2.98\times 10^{45}~{\rm kg}$) & $3h^{-1}~{\rm Mpc}$ & $10$ & $6.78\times 10^{-6}$ & $6.78\times 10^{-4}$ \\
      \hline
      C2: Homogeneity scale / & $\approx (1/6)\pi\rho_{m}D^{3}$ & $300h^{-1}~{\rm Mpc}$ & & & \\
      Supercluster & ($2.98\times 10^{48}~{\rm kg}$) & $30h^{-1}~{\rm Mpc}$ & $10$ & $6.78\times 10^{-4}$ & $6.78\times 10^{-2}$ \\
      \hline
      C3: Hubble sphere / & $\approx (1/6)\pi\rho_{m}D^{3}$ & $6000h^{-1}~{\rm Mpc}$ & & & \\
       --- & ($2.38\times 10^{52}~{\rm kg}$) & --- & --- & $0.27$ & --- \\
      \hline\hline
    \end{tabular}
    \caption{Order--of--magnitude estimates for spatial metric
    and spatial curvature perturbations in six representative
    cases of astrophysical and cosmological gravitating systems
    with two characteristic length scales. For the systems A2 and A3, we
    are considering luminous (baryonic) masses only. The masses in examples
    C1 to C3 were computed from $\rho_{m}$ (baryonic and dark matter)
    given in Eq.~(\ref{wmapdens}) on the assumption that each of them
    is contained within a sphere of radius $D/2$ in flat space. The
    single relevant parameter to characterise the system C3 is the
    ratio between its Schwarzschild radius and its physical radius,
    $\varepsilon$. In this case it does correspond to the matter
    density parameter $\Om_{m}$. The reader is invited to play
    with her/his own set of numbers for $M$, $D$ and $d$ to obtain
    further estimates.}
    \lb{tab:oom1}
\end{table}

The results of our order--of--magnitude estimates are summarised
in Table~\ref{tab:oom1}. It is immediately evident that the two
constraints $\varepsilon \ll 1$ and $\varepsilon(D/d)^{2} \ll 1$,
which would justify the application of a strictly Newtonian or
quasi--Newtonian discription of gravitational interactions, appear
to be simultaneously satisfied only for the cosmological examples
C1 and C2.

However, the cosmic void example C1, which we included
for illustrative purposes, does not really constitute a playground
for perturbation theory in Euclidian space. In this context
(negative) spatial curvature is a zeroth--order effect. In general
relativity, an effective negative spatial curvature is implied for
a void--dominated cosmological model, i.e., a non--flat average
spatial curvature distribution, while the examples above assume a
flat average spatial curvature distribution. The order of
magnitude of spatial curvature is best estimated from the
Gau\ss\ (Hamiltonian, or energy) constraint amongst Einstein's
field equations (see, e.g., Ref.~\ct{buccar2008}), which shows
that the physical contribution by spatial curvature can be
compensated only by a cosmological constant, and this on a single
chosen scale only.

In the homogeneity scale example C2, the order of magnitude of
the spatial curvature perturbation $\varepsilon(D/d)^{2}$ we
find is close to 10\%. Less conservative values for the scale
ratio $D/d$ (with $D$ fixed) would lead to larger values still,
indicating that also in this context perturbation theory as the
sole basis of a dynamical description of structure formation
does not appear to be without its problems of consistency.

Most striking is the violation of the second (curvature)
constraint in the astrophysical systems A2 and A3, bearing
in mind in these cases our restriction to luminous baryonic
matter, and a further, potentially large factor in the mass
content $M$. For such systems, usually, the notion of dark matter
is being invoked in order to account for quantitative deviations
from the Newtonian expectations.

We think that these order--of--magnitude estimates provide a
strong call for a proper relativistic treatment of the underlying
gravitational physics in these systems; spatial curvature is an
inherently relativistic phenomenon, unknown to the Newtonian
theory. The claim on the validity of a quasi--Newtonian metric
according to Eqs.~(\ref{postnewton1})--(\ref{restrictions}) to
describe gravitational physics on {\em all scales\/} in the
observable Universe, apart from black holes and neutron stars, is
thus seriously called into question.

\section{Significance of spatial curvature in cosmology}
We are now going to address an order--of--magnitude estimate for
the spatial curvature in spatially inhomogeneous cosmological
models in the context of a quasi--Newtonian description. However,
we will not look at spatial averages including backreaction effects
(the reader may find such estimates in Ref.~\ct{buccar2008}), but
rather evaluate the spatial Ricci curvature scalar directly from
the quasi--Newtonian spatial metric ansatz of
Eq.~(\ref{postnewton2}), with $\gam_{ij}$ a spatial
metric of constant curvature, i.e., $R(\gam) = \mbox{constant}$.
The Christoffel connection symbols for the
quasi--Newtonian spatial metric are given by
\be
\lb{3chris}
\Gamma_{i}{}^{j}{}_{k}(g)
= \Gamma_{i}{}^{j}{}_{k}(\gam)
- \frac{1}{1-2\Psi}\left[\ptl_{i}\Psi\delta^{j}{}_{k}
+ \ptl_{k}\Psi\delta^{j}{}_{i}
- \ptl_{l}\Psi \gam^{jl}\gam_{ik}\right] \ ,
\ee
while the resultant spatial Ricci curvature scalar
is\footnote{As pointed out in Section~\ref{sec:intro},
neglecting in the quasi--Newtonian framework peculiar
velocities altogether leads to a hypersurface homogeneous
solution of Einstein's field equations, where in the
chosen time slicing the perturbation function $\Psi(t,x^{i})$
is spatially constant so that all spatial gradients of this
function vanish. By a suitable reparametrisation of the time
coordinate $t$, the function $\Psi(t,x^{i})$ can then be set
equal to zero for any arbitrary time interval.}
\be
\lb{qn3rscl}
R(g) = \frac{1}{\ell_{0}^{2}a^{2}(t)(1-2\Psi)}\,\left[\,R(\gam)
+ \frac{4\gam^{ij}D_{i}D_{j}\Psi}{1-2\Psi}
+ \frac{6\gam^{ij}D_{i}\Psi D_{j}\Psi}{(1-2\Psi)^{2}}
\,\right] \ .
\ee
It is standard to introduce the normalisation $R(\gam)=6k$,
where $k\in\{-1,0,+1\}$. Employing the correspondences
\be
\lb{corresp}
\ell_{0}a(t) \leftrightarrow D \ , \quad
\Psi \leftrightarrow \varepsilon \ , \quad\text{and}\quad
D_{i} \leftrightarrow (D/d) \ ,
\ee
we find that
\be
R(g) \approx \frac{1}{D^{2}}\,\left[\,{\cal O}(k)
+ {\cal O}\left(\varepsilon\frac{D^{2}}{d^{2}}\right)
+ {\cal O}\left(\varepsilon^{2}\frac{D^{2}}{d^{2}}\right)
\,\right] \ ,
\ee
or, to first order in $\varepsilon < 1$,
\be
\lb{qn3rscloom}
R(g) \approx \frac{1}{D^{2}}\left[\,k
+ \varepsilon\left(\frac{D}{d}\right)^{2}
+ \ldots\,\right] \ ;
\ee
this result is perfectly in line with J\"urgen's estimate
of Eq.~(\ref{curvest}).

Our next step is to compare the sizes of the perturbation
terms in the quasi--Newtonian spatial Ricci curvature
scalar~(\ref{qn3rscl}), the spatial Laplacian (relative
to $\gam_{ij}$) of the perturbation function $\Psi$ and the
squared spatial gradient of $\Psi$, with each other
on the basis of our estimates for the cosmological
examples C1 and C2 displayed in Table~\ref{tab:oom1}. We
recall that for reasons of simplicity our investigations in
Section~\ref{sec:numex} were grounded on the standard
assumption that the geometry of space is Euclidian and
perturbation theory can be employed to accurately model
deviations thereof. For cases C1 and C2 we now turn to
consider the restrictions of Eq.~(\ref{restrictions}),
which were imposed on spatial metric and spatial curvature
perturbations in Eq.~(2) of Ref.~\ct{ishwal2006}. With
the correspondences of Eq.~(\ref{corresp}), we need to
check whether the inequalities
\be
\lb{restrest}
|\Psi| \ll 1 \ , \quad
|\gam^{ij}D_{i}\Psi D_{j}\Psi|^{2} \ll
|\gam^{ij}D_{i}D_{j}\Psi|^{2}
\quad\Leftrightarrow\quad
|\varepsilon| \ll 1 \ , \quad
\varepsilon^{4}(D/d)^{4} \ll \varepsilon^{2}(D/d)^{4}
\ee
hold. Our results for this consideration are displayed in
Table~\ref{tab:oom2}.
\begin{table}[!htb]
  \centering
    \begin{tabular}{c|c|c|c}
      \hline\hline
      Gravitating system / & & & \\
      Smoothing scale & $\varepsilon$ & $\varepsilon^{4}(D/d)^{4}$ & $\varepsilon^{2}(D/d)^{4}$ \\
      \hline\hline
      C1: Void / & & & \\
      Wall & $6.78\times 10^{-6}$ & $2.11\times 10^{-17}$ & $4.60\times 10^{-7}$ \\
      \hline
      C2: Homogeneity scale / & & & \\
      Supercluster & $6.78\times 10^{-4}$ & $2.11\times 10^{-9}$ & $4.60\times 10^{-3}$ \\
      \hline\hline
    \end{tabular}
    \caption{Comparison of order--of--magnitude estimates for
    spatial metric perturbations $|\Psi|$ and squared spatial
    curvature perturbations $|\gam^{ij}D_{i}\Psi D_{j}\Psi|^{2}$ and
    $|\gam^{ij}D_{i}D_{j}\Psi|^{2}$ for the cosmological cases C1
    and C2 considered in Table~\ref{tab:oom1}.}
    \lb{tab:oom2}
\end{table}

We find that the quasi--Newtonian restrictions according to
Eq.~(\ref{restrest}) are satisfied in both cases. However,
as we argued in Section~\ref{sec:numex}, an exclusively
perturbative approach to modelling the dynamics of structure
formation processes for cosmological gravitating systems
like C1 and C2 is questionable: a large--scale cosmic void
can never be considered just a perturbation of a flat
background space, while for the system C2 the spatial curvature
effect of the matter structure $\varepsilon(D/d)^{2}$ can
easily surpass the 10\% level (cf. our respective estimate of
Section~\ref{sec:numex}) for scale ratios $D/d$ (with $D$ fixed)
only slightly larger than 10. Moreover, as soon as spatial
curvature becomes dynamically significant in the cosmological
context (and, of course, at smaller scales), it is a generic
feature that the local spatial coordinate system used to
describe the dynamics will inevitably break down at a finite
proper distance from the origin. In consequence, a proper
(effective or average) relativistic treatment of the underlying
gravitational interactions appears to be the appropriate one,
implying in particular that the validity of a quasi--Newtonian
spacetime metric, with its associated globally defined coordinate
system, at most scales in the observable Universe is very
doubtful indeed. In contrast to the quasi--Newtonian treatments,
a suitably averaged model can be considered as
a ``background'', and coordinates may be introduced at
this ``background''; but here {\em after\/} effectively smoothing
out local coordinate singularities that may appear on small scales.
We emphasise that local Ricci curvature singularities appearing
in spatially inhomogeneous models are often the result of an
oversimplified matter model such as a hydrodynamical description
of matter, and are not necessarily artifacts of a certain
temporal gauge choice. It ought to be a natural aspiration of
``precision cosmology'' to be aware of and take into account all
of these matters.

Lastly, we emphasise that our geometrical order--of--magnitude
estimates for present--day spatial curvature effects remain
essentially unchanged when employing a temporal gauge alternative
to the Eulerian viewpoint of the longitudinal gauge --- the
simultaneously synchronous and Hubble--flow--comoving (or
matter--comoving) temporal gauge for an irrotational dust fluid
source with 4--velocity field $\vec{u}$, which is also
frequently employed in the description of structure formation
in the late Universe.

\section*{Acknowledgments}
TB and HvE acknowledge kind hospitality by the
Albert--Einstein--Institut, Golm, Germany during a visit in July
2007, and thank Lars Andersson and J\"urgen Ehlers for
stimulating discussions on this occasion.

\addcontentsline{toc}{section}{References}

\end{document}